\documentclass[10pt]{article}
\usepackage{lscape} 

\usepackage{amssymb} 
\usepackage{comment} 

 \def \ccomma{\raise 2pt\hbox{,}} 
 \def \D {\hbox{d}}
 \def \one {``1''}
 \def \two {``2''}

\begin{document}

\title{Solutions of the buoyancy-drag equation 
       with a time-dependent acceleration}

\author{Serge E.~Bouquet${}^{1,2}$, Robert Conte${}^{1,3,4}$, Vincent Kelsch${}^{1}$ and Fabien Louvet${}^{1}$
{}\\
\\ 1. CEA/DAM/DIF, Bruy\`eres-le-Ch\^atel, F--91297 Arpajon Cedex, France
\\ {\hskip 1.0truemm} 
\\ 2. Laboratoire univers et th\'eories (LUTH), Observatoire de Paris, 
\\ Universit\'e de recherche Paris sciences et lettres
    - PSL Research university, 
\\ CNRS, Universit\'e Paris-Diderot, Sorbonne Paris Cit\'e, 
\\ 5, place Jules Janssen, F--92190 Meudon, France
\\ E-mail {Serge.Bouquet@cea.fr}
\\ {\hskip 1.0truemm} 
\\ 3. 
    Centre de math\'ematiques et de leurs applications,
\\ \'Ecole normale sup\'erieure de Cachan, CNRS, 
\\ Universit\'e Paris-Saclay, 
\\ 61, avenue du Pr\'esident Wilson, F--94235 Cachan, France.
\\ {\hskip 1.0truemm} 
\\ 4. Department of mathematics, The University of Hong Kong,
\\ Pokfulam Road, Hong Kong.
\\    E-mail Robert.Conte@cea.fr
}

\maketitle

\hfill 

{\vglue -10.0 truemm}
{\vskip -10.0 truemm}

\begin{abstract}
We perform the analytic study of the 
the buoyancy-drag equation with a time-dependent acceleration $\gamma(t)$
by two methods.
We first determine its equivalence class under the point transformations
of Roger Liouville,
and thus for some values of $\gamma(t)$ define a time-dependent Hamiltonian 
from which the buoyancy-drag equation can be derived.
We then determine the Lie point symmetries of the buoyancy-drag equation,
which only exist for values of $\gamma(t)$ including the previous ones, plus 
additional classes of accelerations for which the equation is reducible to an Abel 
equation. 
This allows us to exhibit two r\'egimes for the asymptotic (large time $t$) 
solution of the buoyancy-drag equation. 
It is shown that they describe a mixing 
zone driven by the Rayleigh--Taylor instability and the Richtmyer--Meshkov 
instability, respectively. 
\end{abstract}



\baselineskip=12truept %



\tableofcontents

\vfill \eject


{2000 Mathematics Subject Classification: 
\\
22E99   (Lie groups) None of the above, but in this section 
\\
34Mxx  Differential equations in the complex domain [See also 30Dxx, 32G34] 
\\
76Fxx  Turbulence [See also 37-XX, 60Gxx, 60Jxx] 
}

http://www.ams.org/msc/msc2010.html


\section{Introduction. The buoyancy-drag equation}

Interface instabilities such as 
Richtmyer--Meshkov instability, 
Kelvin--Helmholtz instability, 
or Rayleigh--Taylor instability, 
are known to produce the mixing of two fluids living on each side 
of a common interface \cite{Chandrasekhar,LlorLNP}.
The Rayleigh--Taylor instability (RTI) plays, however,  
a special role for at least two reasons. 
First, from the physical viewpoint, RTI has been evidenced to occur 
in various objects/processes with a spatial scale ranging from one millimeter 
for inertial confinement fusion laser targets 
to $10^{11}$ meters for supernovae  
and typically $10^{18}$ meters ($\approx 30$ parsec) in supernova remnants. 
The second reason is related to the mathematical description. 
In the linear stage approximation, the formalism of the Richtmyer--Meshkov instability  
can be recovered from the RTI approach where the acceleration $\gamma (t)$ ($t$ is time) 
experienced by the interface is restricted to a kick (impulsive acceleration), 
and, moreover, the dispersion relation of the RTI can be easily derived 
from the Kelvin-Helmholtz instability in a gravitational acceleration $g$  
provided $g$ is formally replaced by $\gamma (t)$ \cite{Chandrasekhar}. 
Because of the broad application of its formalism, 
we have decided in this paper to study the mixing zone produced by the RTI. 
Considering a vertical downward acceleration field, we take a two-fluid configuration 
where the high mass density material (heavy fluid), 
labeled by the subscript \two\, lies above the low mass density medium (light fluid), 
labeled by the subscript \one\ (see \cite{BGP}).
The fluids are assumed to extend 
vertically from $- \infty$ to $+ \infty$ with a velocity of the flow having only a 
vertical component depending on the position in the horizontal plane 
(two-dimensional configuration).
Because of the RTI, spikes of \two\ with length $h_2(t)$ 
(this length is measured from the initial position of the interface) drop into \one\ 
while bubbles of \one\ rise into \two\ with elevation $h_1(t)$ (measured also 
from the initial position of the interface between the two fluids) in the nonlinear regime. 
The mixing zone  
corresponds to a dynamical region that is bounded at its bottom by a 
surface
intersecting the spike tips, and its upper border is given by the 
surface
joining the tops of the bubbles. 
Above (resp.~below) the upper (resp.~lower) frontier, 
the location of which depends upon time, the fluid \two\ (resp.~\one\ ) is pure. 
In a zero-dimensional modeling, 
a spatial average is performed horizontally and
the upper and lower boundaries reduce to two horizontal straight lines with respective 
position $h_1(t)$ above and $h_2(t)$ below the interface and the inner spatial 
structures of the flow cannot be described in between. 
Nevertheless, the height $h(t)$ of the mixing zone 
satisfies $h(t)=h_1(t)+h_2(t)$ and a nonlinear ordinary 
differential equation (ODE) governing the evolution of $h(t)$ can be derived.
This equation is called the buoyancy-drag equation (BDE).
Similarly to an equation of motion, 
the BDE is a second order ordinary differential equation  
that provides the length $h(t)$ as a function of time
%
\begin{eqnarray}
& &
\frac{\D^2 h}{\D t^2}=B \gamma(t) 
 -\frac{C}{h} \left(\frac{\D h}{\D t} \right)^2
\label{eqBDE}
\end{eqnarray}
%
where $B$ (buoyancy coefficient) and $C$ (drag coefficient) are two 
positive constants (see \cite{BGP}), and where $\gamma (t)$ is the 
time-dependent acceleration experienced by the interface.
More precisely, one has $0\le B \le 1$ and $\ 1\le C \le 4$ typically.

Although the BDE was initially 
obtained from phenomenological viewpoints and was considered as an engineering
model describing the motion of the heavy fluid \cite{AHHS}, it has been derived 
by Dimonte \cite{Dimonte} and Dimonte and Schneider \cite{DS00} from the initial 
study by Davies and Taylor \cite{DaviesTaylor}.
This equation has been also derived by Srebro et al. \cite{SESAS} from an 
extension of Layzer approach \cite{Layzer} by Hecht, Alon and Shvarts \cite{HAS94}.
Additionally, a more theoretical approach shows that the BDE comes out from a 
Lagrangian formalism with the inclusion of 
an additional generalized force term to represent the effect of dissipation 
\cite{Ramshaw, Grea}. 
Through this formulation, 
the total energy is preserved and this conservation 
law can be used for estimating global turbulent kinetic energy and dissipation 
rate in the mixing zone. 
\medskip

The motivation of the present paper is to find values of $C$ and, more importantly, $B\gamma(t)$ 
allowing at least a partial integration of (\ref{eqBDE}).
The case of a time-independent (constant) 
acceleration has been studied earlier in 
\cite{ChengGlimmSharp} and \cite{BGP}.
Although only the asymptotic (large value of time) form 
of the solution was found by Cheng, Glimm and Sharp \cite{ChengGlimmSharp}, 
the general solution of the BDE for a constant acceleration has been derived for the first time a 
few years later \cite{BGP}.
In the present work, 
our goal is not to obtain the general solution of (\ref{eqBDE}) for any time-dependent 
acceleration but, instead, to find 
classes of accelerations $\gamma (t)$ for which the BDE can be 
fully or partially integrated.
\medskip

There exist three general methods to investigate a given second order 
ordinary differential equation. 
The first one corresponds to the Painlev\'e analysis \cite{CMBook}. 
It is based on those singularities of the general solution which depend on the initial conditions. 
The second approach applies to the class of equations, 
to which the BDE belongs, 
investigated by Roger Liouville \cite{RLiouvilleODE2},
and in the third method the differential order of the BDE is reduced by 
using Lie symmetries.

In our case, the first method cannot apply, 
since a necessary condition of its applicability
is the existence of a relative integer $n$ such that 
$C=-1+1/n$ \cite{CMBook}, but this condition never happens.

However a useful property is the form invariance of equation (\ref{eqBDE}) \cite{Louvet}.
Indeed, the one-parameter point transformation $\varphi_a$, 
%
\begin{eqnarray}
& & 
T=\frac{a t}{t+a},\
h=\left(\frac{t}{T}\right)^{1/(1+C)} H,\
\gamma= \left(\frac{t}{T}\right)^{-(3+4C)/(1+C)} \Gamma,\
a \hbox{ arbitrary},
\label{eqForm_invariance} 
\end{eqnarray}
%
only changes the acceleration,
%
%
\begin{eqnarray}
& &
\frac{\D^2 H}{\D T^2}=B\ \Gamma(T) 
-\frac{C}{H} \left(\frac{\D H}{\D T} \right)^2.
\label{eqBDEcaps}
\end{eqnarray}
%

The inverse of $\varphi_{a}$ is $\varphi_{-a}$ 
and the composition of two such transformations does not 
generate a new transformation,
since $\varphi_b \varphi_a=\varphi_{a b /(a+b)}$.
In particular, a power-law accelation $\gamma \sim t^n$
is mapped to another power-law $\gamma \sim t^{-\alpha-n}$,
with
%
\begin{eqnarray}
& &
\alpha=\frac{3+4 C}{1+C}=4-\frac{1}{1+C}\ccomma
\label{EqDefAlpha}
\end{eqnarray}
%
the fixed point of such a map being $\gamma \sim t^{-\alpha/2}$,
which will indeed be encountered below.

The paper is organized as follows.
In section \ref{sectionExisting_solutions},
we first recall the existing analytic solutions of the BDE.
In section \ref{sectionLiouville}, 
we build a Hamiltonian description of the dissipative equation (\ref{eqBDE})
and, for specific $t$-dependences of the acceleration $\gamma(t)$,
find an invariant of the resulting Hamiltonian system.
In section \ref{sectionLie_point_symmetries},
for selected accelerations $\gamma(t)$,
we lower the differential order by one unit.

%
%

\section{Existing solutions}
\label{sectionExisting_solutions}

Some solutions of Eq.~(\ref{eqBDE}) for the RTI are already known in four cases,
the first three have been derived using Lie symmetries
 \cite{BGP,Louvet,Kelsch}.
Throughout the paper, $t_0$ denotes some time unit.

\begin{enumerate}
\item
For a constant acceleration $\gamma = \gamma_0$, 
the BDE is autonomous and admits the first integral 
%
\begin{eqnarray}
& & \gamma = \gamma_0:\
K=h^ {2 C} \left\lbrack
  \left(\frac{\D h}{\D t} \right)^2-\frac{2 B \gamma_0}{1+2 C} h\right\rbrack \ .
\label{eq-gamma-constant} 
\end{eqnarray}
%
The BDE is then invariant under the two symmetries
$\partial_t$ and $t \partial_t + 2 h \partial_h$ (the subscripts represent the partial derivatives),
which allows us to obtain its general solution in the implicit form \cite{BGP}
\cite[formula 3.194]{GRBook}
%
\begin{eqnarray}
& &  {\hskip -15.0truemm}
\gamma = \gamma_0,\
t=t_1+\frac{h^{1+C}}{(1+C)\sqrt {\varepsilon K}}  
\ {}_2 F_1\left (\frac{1}{2}, \frac{1+C}{1+2C},\frac{2+3C}{1+2C},
-\frac{2B \gamma_0 h^{1+2C}}{(1+2C) \varepsilon K } \right), \ 
\varepsilon=\hbox{sign}(K),
\label{eqSol_gamma_constant} 
\end{eqnarray}
%
in which ${}_2 F_1$ is the hypergeometric function of Gauss,
and the two arbitrary constants are $t_1$ and $K$.

This expression yields the asymptotic behavior
of the height as $t \to +\infty$
%
\begin{eqnarray}
& &
h \sim \frac{B \gamma_0}{2(1+2C)} \ t^2 \ .
\label{eqSol_asymp_gamma_constant} 
\end{eqnarray}
%
%
Such a time-dependence has already been found earlier by Neuvazhaev \cite{N75}.

This quadratic law is consistent with the fact that
in a constant acceleration field the free fall distance scales as $t^2$.
This result is not really surprising because the friction term 
$- (C/h)(\D h/\D t)^2$ scales as $t^0$ and the BDE becomes a free fall equation.

In a constant acceleration field $\gamma_0$
without friction ($C=0$) ,
the general solution of 
Eq.~(\ref{eqBDE}) behaves like $h \sim (B \gamma_0 / 2) \ t^2$,
and the comparison with 
(\ref{eqSol_asymp_gamma_constant}) shows, 
however,
that the actual solution accounts for friction effects. 
We could have expected a balance between the 
acceleration term $B \gamma_0$ and the friction term $ - (C/h)(\D h/\D t)^2$, 
i.e. $h \sim [B \gamma_0 /(4 C)] \ t^2 $, in the asymptotic 
evolution. Nevertheless, Eq.~(\ref{eqSol_asymp_gamma_constant}) 
proves that this condition is not fulfilled and, instead, the balance 
between the three terms (acceleration, friction and inertia $\D^2 h/\D t^2$) 
is achieved in the BDE.

\item
For the power law acceleration $\gamma \sim t^{-\alpha}$,
the general solution is simply obtained from (\ref{eqSol_gamma_constant})
by action of the transformation (\ref{eqForm_invariance})
to generate the case $\Gamma=\gamma_0$ in (\ref{eqBDEcaps}). 

As expected, for large time $t$, the behavior $h \sim t^2$ is no longer valid. 
The actual asymptotic solution is obtained by noting that $\lim_{t \to +\infty} T =a$.
As a consequence, $T$ and $H$ are bounded with $H \to H(a)$. 
Using (\ref{eqForm_invariance}), we get $h \sim (t/a)^{1/(1+C)} H(a)$, \textit{i.e.}
%
\begin{eqnarray}
& &
h \sim t^{1/(1 + C)},\ t \to +\infty.
\label{eqSol_asymp_power_alpha} 
\end{eqnarray}

This behavior is actually 
very surprising for the Rayleigh--Taylor mixing since such a power law is 
relevant to the Richtmyer--Meshkov instability (no acceleration or impulsive 
acceleration when a shock wave front reaches an interface). 
Indeed, for a zero acceleration in (\ref{eqBDE}), the BDE reduces to 
%
\begin{eqnarray}
& &
\frac{\D^2 h}{\D t^2}= -\frac{C}{h} \left(\frac{\D h}{\D t} \right)^2,
\label{eqBDE-RMI}
\end{eqnarray}
%
and the general solution is 
%
\begin{eqnarray}
& & \gamma = 0:\
h_{\rm RM} = K (t - t_1)^{1/(1+C)}
\label{eqSolBDE-RMI} 
\end{eqnarray}
%
where the subscript RM stands for Richtmyer--Meshkov and where  $K$ and $t_1$ 
are the two constants of integration. 
Asymptotically (large time $t$), we get
%
\begin{eqnarray}
& &
h_{\rm RM} \sim t^{1/(1 + C)},\ t \to +\infty.
\label{eqSolBDE-RMI-asymp} 
\end{eqnarray}
%
As a consequence, the physical interpretation of (\ref{eqSol_asymp_power_alpha}) is as 
follows: 
for $n=- \alpha$, the acceleration $\gamma (t)$ decreases very quickly 
with time and the mixing regime is not driven by the RTI at large time. 
Instead, the system experiences the Richtmyer--Meshkov instability (RMI)
asymptotically because $\gamma \to 0$ for $t \to +\infty$.

Finally, as in the linear stage of the RMI we have a ballistic growth of the 
mixing zone, we get $h \sim t$.  
However, (\ref{eqSol_asymp_power_alpha}) shows that 
in the nonlinear regime, $h$ increases more slowly than linearly in $t$ 
(the exponent $1/(1+C)$ is always smaller than $1$) and the observed 
deceleration is produced by the friction term $- (C/h)(\D h/\D t)^2$.

\item
For a general power law acceleration $\gamma \sim t^n$ 
with an arbitrary real exponent $n$, 
the BDE (\ref{eqBDE}) only admits the symmetry $t \partial_t + (2+n) h \partial_h$,
which allows its reduction
to an Abel equation \cite{BGP}, 
%
\begin{eqnarray}
& &
\left\lbrace
\begin{array}{ll}
\displaystyle{
\gamma=\gamma_0 \left(\frac{t}{t_0}\right)^n,\ 
I=\left(\frac{t}{t_0}\right)^{-n-2} h,\
J=(1+C) \left(\frac{t \D h}{h \D t} -n-2\right),\
}\\ \displaystyle{
I J \D J  +(1+C)
\left[\left(J+(1+C)n+\frac{3+4 C}{2}\right)^2 - \frac{1}{4} -(1+C) \frac{B \gamma_0 t_0^{2}}{I}\right] \D I=0.
}
\end{array}
\right.
\label{eqPoint-sym-t0} 
\end{eqnarray}
%
Since this Abel ODE is in general not integrable 
\textbf{
\cite{ChebTerrab-Roche,Murphy,Pana2011,PolyaninZaitsev},
}
this only defines the zero-parameter scaling solution,
%
\begin{eqnarray}
& &
\left\lbrace
\begin{array}{ll}
\displaystyle{
\gamma=\gamma_0 \left(\frac{t}{t_0}\right)^n,\ 
}\\ \displaystyle{
h=h_0 \left(\frac{t}{t_0}\right)^{n+2},\ 
h_0=\frac{B \gamma_0 t_0^2}{(n+2) [n (1 + C) + 1 + 2 C]} ,\
(n+2)\left(n+\frac{1+2 C}{1+C}\right) \not=0.
}
\end{array}
\right.
\label{eqSol-tpowern} 
\end{eqnarray}
%
In these relations, the exponent $n$ must satisfy two major constraints. 
First, as the mixing zone thickness increases with time, the exponent $n$ 
should obey $n + 2 > 0$.
Second, the coefficient $h_0$ has to be positive and, since $n > - 2$, 
we get
%
\begin{eqnarray}
& &
n > - \frac{1 + 2C}{1 + C} = - 2 + \frac{1}{1 + C} \ .
\label{eqExposant-n-limite}
\end{eqnarray}
%
As a consequence, one expects the solution $h \sim t^{n+2}$ to be valid for $n$  
either positive or negative, provided its value is larger than the lower bound given 
by Eq.~(\ref{eqExposant-n-limite}). 
For $n>0$, the acceleration grows with time and the thickness of the mixing zone 
increases. For the case $n<0$, although the acceleration vanishes for large time, 
the thickness of the mixing zone increases too. In both cases, the mixing is driven 
by the RTI. 

Moreover, for the lower bound (\ref{eqExposant-n-limite}), 
we have $n + 2 = 1/(1 + C)$ and the time dependence in 
(\ref{eqSol_asymp_power_alpha}) and (\ref{eqSol-tpowern}) coincide.
According to (\ref{eqSolBDE-RMI-asymp}), this specific time variation 
of the acceleration produces the same asymptotic mixing rate for the RTI than 
for the RMI.
Finally, for $n = 0$ we have $h_0 = B \gamma_0 t_0^2 /[2 (1 + 2C)]$ 
and therefore $h = B \gamma_0 t^2 / [2 (1 + 2C)] $. 
This solution corresponds to (\ref{eqSol_asymp_gamma_constant}) 
and we conclude that the particular solution (\ref{eqSol-tpowern}) 
provides the correct asymptotic behavior of the RTI for a constant 
acceleration.

If the acceleration decreases with $t$ faster than prescribed by the lower bound value 
(\ref{eqExposant-n-limite}), the solution $h \sim t^{n+2}$ does not happen anymore. 
According to our explanations, we might expect that the asymptotic solution 
of the BDE be given by (\ref{eqSolBDE-RMI-asymp}).
This prediction actually arises for $\gamma \sim t^{- \alpha}$ [see item (2) above]: 
the value $n = - \alpha = - 2 - [(1 + 2C)/(1 + C)]$
is always smaller than $- 2$ and is therefore below the lower value 
(\ref{eqExposant-n-limite}), and Eq.~(\ref{eqSol_asymp_power_alpha}) shows 
that $h(t)$ behaves like $t^{1/(1 + C)}$ instead of $t^{n + 2} = t^{- (1 + 2C)/(1 + C)}$ 
that decreases with time.

In the next item, this prediction is also proven to hold for another negative 
value of $n$.


\item
For the value of $n$ which leaves the behaviour $\gamma \sim t^n$
invariant under the map $\varphi_a$,
i.e.~$n=-\alpha/2$,
the Abel equation (\ref{eqPoint-sym-t0})
becomes linear in $J^2$
and there exists a first integral
%
\begin{eqnarray}
& &
n = - \frac{\alpha}{2},\
\gamma = \gamma_0 \left(\frac{t}{t_0}\right)^n,\
K=\left\lbrack J^2 -\frac{1}{4}
-2 \frac{(1+C)^2}{1+2 C} \frac{B \gamma_0 t_0^{2}}{I}
  \right\rbrack I^{2(1+C)}.
\label{eq-gamma-power-half_alpha} 
\end{eqnarray}
%
Moreover, when $K$ vanishes,
the first order ODE for $h(t)$ can be integrated.
Indeed, the change of function 
%
\begin{eqnarray}
& &
(h,t) \to (H,T):\ h=T H,\ T=\left(\frac{t}{t_0}\right)^{n+2},
\end{eqnarray}
%
maps the first integral to
%
\begin{eqnarray}
& &
T^2 \left(\frac{\D H}{\D T}\right)^{2}- H (H-H_0)=0,\ 
H_0= - 8 B \gamma_0 t_0^2 \frac{(1+C)^2}{1+2 C}\ccomma
\end{eqnarray}
%
an equation linearizable by derivation.
This ODE has two kinds of solutions \cite{ChazyThese}:
the so-called singular solution $H=H_0$,
$h=H_0 (t/t_0)^{2-\alpha/2}$,
which must be rejected because it is never solution of the BDE,
and the general solution
\cite{Louvet}, 
%
\begin{eqnarray}
& &
H=(c_1 T^{-1/2}+c_2 T^{1/2})^2,\
c_1 c_2=\frac{H_0}{4}\ccomma\ \frac{c_1}{c_2} \hbox{ arbitrary},
\end{eqnarray}
%
\textit{i.e.}~
\begin{eqnarray}
& & {\hskip -10.0truemm}
n = - \frac{\alpha}{2},\
\gamma=\gamma_0 \left(\frac{t}{t_0}\right)^n,\ 
h=
\left\lbrack c_1 + c_2 \left(\frac{t}{t_0}\right)^{n+2}\right\rbrack^2,\
c_1 c_2=\frac{H_0}{4}\cdot 
\label{eqSolpower-alpha_over-two}
\end{eqnarray}
%
The latter solution depends on the arbitrary parameter $c_1/c_2$
but it never reduces to the previous solution (\ref{eqSol-tpowern}).\

The exponent $n+2$ in (\ref{eqSolpower-alpha_over-two}) is always positive 
and hence, for large time $t$, the constant $c_1$ can be neglected in 
(\ref{eqSolpower-alpha_over-two}) showing that  $h$ grows like $t^{2(n+2)}$,
which is,
%
\begin{eqnarray}
& &
h \sim t^{1/(1 + C)}.
\label{eqSol_asymp_power_alpha/2} 
\end{eqnarray}
%
Interestingly, the behavior (\ref{eqSol_asymp_power_alpha}) is recovered. 

This result is not surprising because the exponent $n= - \alpha /2$,
which can be written as the sum $n= - 2 + 1/(1 + C) - 1 / [2(1 + C)]$
is therefore smaller than the lower bound  (\ref{eqExposant-n-limite}).

We conclude that, although the acceleration 
decreases with $t$ more slowly for $n = - \alpha /2$ than for $n = - \alpha $, 
the function $\gamma (t)$ vanishes rapidly enough for the system to exhibit a 
mixing zone driven by the RMI instead of the RTI.

\end{enumerate}
%

\section{Method of Roger Liouville for second order equations}
\label{sectionLiouville}

The approach developed by 
R.~Liouville  \cite{RLiouvilleODE2}
applies to the class of equations 
%
\begin{eqnarray}
& &
{\hskip -15.0truemm}
\frac{\D^2 h}{\D t^2}
+  a_3(h,t) \left(\frac{\D h}{\D t} \right)^3
+3 a_2(h,t) \left(\frac{\D h}{\D t} \right)^2
+3 a_1(h,t)       \frac{\D h}{\D t}
+  a_0(h,t)=0,
\label{eqOrder2ClassLiouville}
\end{eqnarray}
%
whose property is to be form invariant under the point transformation
%
\begin{eqnarray}
& &
{\hskip -15.0truemm}
(t,h) \to (T,H):\
t=F(T,H),\ h=G(T,H).
\label{eqTransfo-Order2}
\end{eqnarray}
%
By determining the invariants of (\ref{eqOrder2ClassLiouville}) 
under the transformation (\ref{eqTransfo-Order2}),
one may be able to integrate.
A nice account of this method can be found in \cite{BB1999},
to which we refer for the notation.
In the case of (\ref{eqBDE}), one obtains
%
\begin{eqnarray}
& &
L_1=C B \gamma(t) h^{-2},\
L_2=0,\
\nu_5=0,\
w_1=0,\
i_2=C(2+C) B \gamma(t) h^{-3},\ 
\end{eqnarray}
%
therefore (see e.g.~\cite[Lemma 1 page 458]{BB1999})
the ordinary differential equation (\ref{eqBDE}) can be mapped to an ODE (\ref{eqOrder2ClassLiouville})
in which $a_3=a_2=a_1=0$, 
%
\begin{eqnarray}
& &
t=t,\ h=H^{1/(1+C)},\
\frac{\D^2 H}{\D t^2}-  (1+C) B \gamma(t) H^{C/(1+C)}=0. 
\label{eqBDEHamilton}
\end{eqnarray}

Since this equation is independent of $\D H/ \D t$, it can be interpreted as the 
Hamilton equation of a classical time-dependent Hamiltonian $\mathbb{H}$ 
defined by
%
\begin{eqnarray}
& &
\mathbb{H}(q,p,t)=\frac{p^2}{2} + V(q,t),\ 
q=H,\
p=\frac{\D q}{\D t},\ 
V=-\frac{(1+C)^2}{1+2 C} B \gamma(t) q^\frac{1+2 C}{1+C} \  
\end{eqnarray}
%
where $q$, $p$ and $V$ are respectively the position, linear momentum and potential.
Under this transformation, equation~(\ref{eqBDEHamilton}) becomes a standard equation 
of motion 
%
\begin{eqnarray}
& &
\frac{\D^2 \ q}{\D t^2} \ + \  \frac{\partial}{\partial q} \ V(q,t) = 0,
\end{eqnarray}
%
for which we are going to look for an invariant of motion $\mathbb{I}(q,p,t)$.

The invariant obeys the equation 
$ \partial \mathbb{I} / \partial t + p (\partial \mathbb{I} / \partial q) - (\partial V / \partial q) (\partial \mathbb{I} / \partial p) = 0$
and since $\mathbb{H}$ is quadratic in $p$, 
it is natural to look for $\mathbb{I}(q,p,t)$ also quadratic in $p$.
Such an invariant does exist, 
%
\begin{eqnarray}
& &
\mathbb{I}(q,p,t)=(d_1 t^2 + 2 d_2 t + d_3)  \mathbb{H}(q,p,t)
 -(d_1 t + d_2) q p
 +d_1 \frac{q^2}{2}\ccomma\
(d_1,d_2,d_3) \hbox{ arbitrary},
\end{eqnarray}
%
however only for specific accelerations $\gamma(t)$ defined by,
%
\begin{eqnarray}
& & {\hskip -10.0truemm}
\left(d_1 t^2 + 2 d_2 t+d_3 \right) \frac{\D \gamma}{\D t}
 +(d_1 t + d_2) \alpha \gamma=0,\
 \alpha=\frac{3+4 C}{1+C}\cdot
\label{eqODEgammaHamiltonian}
\end{eqnarray} 
%
According to the values of the $d_i$'s, three types of $t$-dependence for $\gamma(t)$ 
come out,
%
\begin{eqnarray}
& & \gamma=
\left\lbrace
\begin{array}{ll}
\displaystyle{
\gamma_0 \left(\frac{t^2-t_1^2}{t_0^2}\right)^{-\alpha/2} \ \ (d_1 \not=0),
}\\ \displaystyle{
\gamma_0 \left(\frac{t}{t_0}\right)^{-\alpha/2}\ (d_1=0,\ d_2\not=0),
}\\ \displaystyle{
\gamma_0\  \ (d_1=d_2=0,\ d_3\not=0),
}
\end{array}
\right.
\label{eqgamma-Hamilton}
\end{eqnarray}
%
where $t_1^2$ is an arbitrary real constant of any sign.

The first expression in (\ref{eqgamma-Hamilton}) has been obtained 
by performing 
a time translation $t \to t - d_2/(2 d_1)$ 
(which leaves the BDE invariant),
and
$t_1^2 = (d_2)^2 / (4 d_1^2) - d_3/d_1$.

This expression is an extension of the second 
case listed in section \ref{sectionExisting_solutions} 
and it is recovered for $t_1^2 = 0$. 

The last two values correspond to the first and fourth 
cases listed in section \ref{sectionExisting_solutions}
(the second value requires the shift $t \to t - d_3/d_2$).

Le us now show that one can recover these three values by another approach,
and even find more general ones.

 \vspace{0.2cm}
   

\section{Method of Lie point symmetries}
\label{sectionLie_point_symmetries}

Given any partial differential equation $E(x,t,u(x,t), u_x, u_t, \dots)=0$,
a Lie point symmetry is a transformation
\begin{eqnarray}
& &
{\hskip -15.0truemm}
(x,t,u) \to (X,T,U):\
X=F(x,t,u),\ T=G(x,t,u),\ U=H(x,t,u),\ 
\nonumber 
\end{eqnarray}
mapping a solution $u(x,t)$ to another solution $U(X,T)$.

Practically, instead of this finite transformation,  
one computes the infinitesimal transformation
%
\begin{eqnarray}
& &
{\hskip -12.0truemm}
X=x+\varepsilon \zeta(x,t,u),\
T=t+\varepsilon      \xi(x,t,u),\
U=u+\varepsilon  \eta(x,t,u),
\label{Lie-infinitesimal}
\end{eqnarray}
%
associated to the infinitesimal point symmetry 
$S=\zeta \partial_x +\xi \partial_t +\eta \partial_u$ where the subscripts stand for the 
partial derivatives \cite{OlverBook,OvsiannikovBook}.

In the case of (\ref{eqBDE}), the assumption 
%
\begin{eqnarray}
& &
T=t+\varepsilon  \xi(t,h),\
H=h+\varepsilon \eta(t,h)
\label{eqPoint_Lie_ht}
\end{eqnarray}
%
yields the set of determining equations for $\xi(t,h),\eta(t,h)$
\cite{Kelsch} 
%
\begin{eqnarray}
& &
\left\lbrace 
\begin{array}{ll}
\displaystyle{
\xi_{hh} - (C/h) \xi_h=0,
}\\ \displaystyle{
\eta_{hh} + (C/h) \eta_h -(C/h^2) \eta -2 \xi_{th}=0,
}\\ \displaystyle{
\xi_{tt} + 3 B \gamma(t) \xi_h -2 (C/h) \eta_t -2 \eta_{th}=0,
}\\ \displaystyle{
\eta_{tt} - 2 B \gamma(t) \xi_t + B \gamma(t) \eta_h - B \gamma'(t) \xi=0.
}
\end{array}
\right.
\label{eqPoint_determining} 
\end{eqnarray} 
%
where $\gamma'(t)$ is the time derivative of $\gamma(t).$

\textit{Remark}.
The above assumption (\ref{eqPoint_Lie_ht}) for Lie point symmetries 
does not act on $\gamma(t)$,
therefore it cannot detect the form invariance (\ref{eqForm_invariance}).

The system (\ref{eqPoint_determining}) 
is a linear overdetermined system for $\xi(t,h),\eta(t,h)$,
solved as follows.

The first equation is an ODE for $\xi(h)$ (with $t$ a parameter) having the type of Fuchs, 
whose general solution depends on two arbitrary functions of $t$,
%
\begin{eqnarray}
& &
\xi=C_1(t) \ \frac{h^{1+C}}{1+C} + C_2(t).
\label{eq-xi}
\end{eqnarray}
%

The second equation is then an ODE for $\eta(h)$ (with $t$ a parameter) of the type of Fuchs, 
whose general solution introduces two more arbitrary functions of $t$,
%
\begin{eqnarray}
& &
\eta=C_3(t) h + C_4(t) h^{-C} + C_1'(t) \frac{h^{2+C}}{(1+C)^2}
\label{eq-eta}
\end{eqnarray}
%
where $C_1'(t)$ stands for the time derivative of $C_1(t)$.

Inserting these two expressions in the last two equations 
(\ref{eqPoint_determining}) puts constraints on $C_j(t)$ and $\gamma(t)$.

Finally, $\xi$ and $\eta$ depend on four arbitrary constants $d_i$ 
\cite{Kelsch}
\begin{eqnarray}
& &
 \xi=d_1 t^2 + 2 d_2 t +d_3,\
\eta = \frac{d_1 t + d_2 + d_4}{1+C} h,
\nonumber 
\end{eqnarray} 
%
and $\gamma(t)$ must obey a first order linear ODE,
%
\begin{eqnarray}
& & {\hskip -10.0truemm}
\left(d_1 t^2 + 2 d_2 t+d_3 \right) \frac{\D \gamma}{\D t}
 +\left[(d_1 t + d_2 + d_4) \alpha- 4 d_4 \right] \gamma=0
\label{eqODEgammaSymmetries}
\end{eqnarray} 
%
whose only fixed parameter is the positive constant $\alpha$. 
This ODE is an extension of (\ref{eqODEgammaHamiltonian}). 

This four-dimensional algebra, generated by
%
\begin{eqnarray}
& &
X_1= t^2\partial_t + t h \partial_h,\
X_2= 2 t\partial_t +   h \partial_h,\
X_3=    \partial_t,
X_4=                   h \partial_h,\
\end{eqnarray} 
%
is decomposable into
$\left\lbrace X_1, X_2, X_3\right\rbrace \oplus X_4$
and its nonzero commutators are, 
%
\begin{eqnarray}
& &  {\hskip -10.0truemm}
[X_1,X_2]=-2 X_1,\ 
[X_1,X_3]=-X_2,\
[X_2,X_3]=-2 X_3.
\end{eqnarray}
%




Since the ODE (\ref{eqODEgammaSymmetries}) contains one more parameter than
the similar ODE (\ref{eqODEgammaHamiltonian}) resulting from the Hamiltonian structure,
the set of values of $\gamma(t)$ is now larger, 
%
\begin{eqnarray}
& & \gamma=
\left\lbrace
\begin{array}{ll}
\displaystyle{
\gamma_0 \left(\frac{t-t_1}{t_0}\right)^{n_1}
\left(\frac{t+t_1}{t_0}\right)^{n_2},
\ n_1+n_2=-\alpha,\ n_1-n_2=\alpha \frac{t_2}{t_1}\ \ (d_1 \not=0),
}\\ \displaystyle{
\gamma_0 \left(\frac{t}{t_0}\right)^{n}\ (d_1=0,\ d_2\not=0),
}\\ \displaystyle{
\gamma_0 e^{\alpha t/t_3}\  \ (d_1=d_2=0,\ d_3\not=0, \ d_4\not=0), 
}\\ \displaystyle{
\gamma_0\ (d_1=d_2=d_4=0,\ d_3\not=0),
}
\end{array}
\right.
\label{eqgamma_Lie} 
\end{eqnarray}
%
in which $t_1$, $t_2$, $t_3$ and $n$ are arbitrary real constants,
and time has been shifted like in (\ref{eqgamma-Hamilton})
in order to derive the first three expressions.

The second and fourth values of $\gamma(t)$ 
yield the solutions already mentioned 
in section \ref{sectionExisting_solutions},
 but the first and the third ones are new.

The first value defines a new case of reduction
to an Abel equation in which the invariants $I$ and $J$ are given by
%
\begin{eqnarray}
& &
\left\lbrace
\begin{array}{ll}
\displaystyle{
\frac{\D \gamma}{\gamma \D t}=-\alpha \frac{t-t_2}{t^2-t_1^2},\
t_2 \hbox{ and } t_1=\hbox{arbitrary real constants},\ 
}\\ \displaystyle{
\gamma=\gamma_0 \left(\frac{t-t_1}{t_0}\right)^{n_1}
                \left(\frac{t+t_1}{t_0}\right)^{n_2},\
n_1+n_2=-\alpha,\ n_1-n_2=\alpha t_2/t_1,\ 
}\\ \displaystyle{
I=\left(\frac{t-t_1}{t_0}\right)^{ - n_1 - 2}
  \left(\frac{t+t_1}{t_0}\right)^{ - n_2 - 2} h,
}\\ \displaystyle{
J=\frac{1}{t_0} \left\lbrack(1+C) (t^2-t_1^2) \frac{\D h}{h \D t} -t-(3+4 C) t_2\right\rbrack ,
}\\ \displaystyle{
I J \frac{\D J}{\D I}+(1+C)
\left[ \left( J+(3+4 C) \frac{t_2}{t_0} \right)^2 - \left( \frac{t_1}{t_0} \right)^2 -(1+C) \frac{B \gamma_0 t_0^2}{I}  \right] =0.
}
\end{array}
\right.
\label{eqPoint-sym-t1t2} 
\end{eqnarray}
%
This solution contains the first expression of the list (\ref{eqgamma-Hamilton}) as a special case 
for $n_1=n_2=-\alpha /2$.

The only closed form solution which this defines is,
%
\begin{eqnarray}
& &  {\hskip -10.0truemm}
h = \frac{(1+C)B \gamma_0 t_0^4}{(3+4 C)^2 \ t_2^2 \ - \ t_1^2}
  \left(\frac{t-t_1}{t_0}\right)^{n_1+2}
  \left(\frac{t+t_1}{t_0}\right)^{n_2+2},  
\label{eqSolPoint-sym-tn}
\end{eqnarray}
and it depends on the two arbitrary constants $t_1$, $t_2$.

Let us examine the asymptotic form of this solution.
For large time ($t \gg t_1$), the acceleration and the height are 
%
\begin{eqnarray}
& &
\gamma \sim t^{n_1+n_2}, \ \ h \sim t^{n_1+n_2+4},\ \ t \to +\infty.
\label{eqSol_asympNew1} 
\end{eqnarray}
%
Obviously, this behavior does not belong to the class (\ref{eqSol-tpowern}) with $n=n_1+n_2$. 
However, with (\ref{EqDefAlpha}) and the constraint $n_1+n_2=-\alpha$, we obtain 
$h \sim t^{1/(1+C)}$ which not surprisingly is identical to (\ref{eqSol_asymp_power_alpha}). 
The RMI growth is again recovered.

Similarly, the third value in the list (\ref{eqgamma_Lie})
also defines a reduction to an Abel equation 
%
\begin{eqnarray}
& &
\left\lbrace
\begin{array}{ll}
\displaystyle{
\gamma=\gamma_0 e^{\alpha t/t_3}, ~ t_3 \not= 0
}\\ \displaystyle{
I=e^{-\alpha t/t_3}h,
}\\ \displaystyle{
J=(1+C) \left( t_3 \frac{\D h}{h \D t} - \alpha \right),
}\\ \displaystyle{
I J \frac{\D J}{\D I}+(1+C)
 \left[ \left( J+3+4 C \right)^2 -(1+C) \frac{B \gamma_0  t_3^2}{I}  \right]=0
}
\end{array}
\right.
\label{eqPoint-sym-t1t2exp} 
\end{eqnarray}
%
and the particular solution
%
\begin{eqnarray}
& & {\hskip -8.0truemm}
h=\frac{B \gamma_0 t_3^2}{\alpha^2 (1+C)} e^{\alpha t/t_3}.
\label{eqBDE-exponential-solution}
\end{eqnarray}
%
It is clear 
that this solution arises only for $t_3>0$: if $t_3<0$, the acceleration 
decreases exponentially 
with time and after a transient phase, $h$ will follow the RMI law
(\ref{eqSol_asymp_power_alpha}) instead of (\ref{eqBDE-exponential-solution}).

The three ODEs for $J(I)$ defined in (\ref{eqPoint-sym-t0}), (\ref{eqPoint-sym-t1t2}) 
and (\ref{eqPoint-sym-t1t2exp}) 
are Abel equations of the second kind,
%
\begin{eqnarray}
& &
I J \frac{\D J}{\D I} + a \left[(J-b)^2 - c^2 -\frac{d}{I}  \right] =0,
\label{eqAbel-IJ}
\end{eqnarray}
%
for the respective values
%
\begin{eqnarray}
& &
a=1+C,\ d=(1+C) B \gamma_0 t_0^2,\ c=\frac{1}{2},\ b=-(1+C) n -\frac{3+4 C}{2},
\\ & &
a=1+C,\ d=(1+C) B \gamma_0 t_0^2,\ c=\frac{t_1}{t_0},\ b=-(3+4 C) \frac{t_2}{t_0},
\\ & &
a=1+C,\ d=(1+C) B \gamma_0 t_3^2,\ c=0,\               b=-(3+4 C) \cdot
\end{eqnarray}
%

The classical method to investigate the integrability of an Abel ODE is recalled
in the \ref{sectionAbel}.
In our case (\ref{eqAbel-IJ}), the first two relative invariants of Roger Liouville evaluate to
%
\begin{eqnarray}
& & {\hskip -8.0truemm}
s_3=\frac{2 a^2 b}{27 I^4}[a(b^2-9 c^2) I - 9 d (a-1)], 
\nonumber\\ & & {\hskip -8.0truemm}
s_5=\frac{2 a^3 b}{27 I^7} \left[ 
a^2(b^2-9 c^2) (b^2+3 c^2) I^2 
-3 d (a-1) [(2 b^2 a+18 c^2 a-3 b^2+3 c^2) I + 3 d (3 a-2)]
\right].
\label{eqInvariants-sm}
\end{eqnarray}
%

Since the absolute invariant $s_5^3/s_3^5$ is generically nonconstant, 
there exists no change of variables making Eq.~(\ref{eqAbel-IJ}) separable.

The conditions $s_3=s_5=0$ define two nongeneric cases, 
$b=0$ and $(a=1,b^2=9c^2)$, 
the second one being nonphysical. 
Moreover,
the Maple package developed in \cite{ChebTerrab-Roche},
which knows all the Abel ODEs which have been integrated before the year 2000,
fails to map (\ref{eqAbel-IJ}) to one of the integrated equivalence classes,
except in the unphysical case $d=0$.\\

To summarize,
the only cases when a first integral $K$ is known to Eq.~(\ref{eqAbel-IJ}) 
are,
%
%
%
%
%
%
%
\begin{eqnarray}
& & {\hskip -8.0truemm}
\left\lbrace
\begin{array}{ll}
\displaystyle{
b=0, a\not=1/2:\ K=\log I+\frac{\log[(2a-1)(J^2-c^2)-2 a d/I]}{2 a},
}\\ \displaystyle{
b=0, a=1/2:\ K=\log I+\frac{(J^2-c^2)I}{d},
}\\ \displaystyle{
a=1,c^2\not=b^2:\ K=(c+b) \log[d+(b-c) I (J-b-c)]+(c-b)\log[d+(b+c) I (J-b+c)],
}\\ \displaystyle{
a=1,c^2=b^2:\ K=\log[2 b I J+d]-2 \frac{b}{d} I (J-2 b),
}\\ \displaystyle{
d=0, c\not=0:\ K=\log I+\frac{(c+b)\log[J-b-c]+(c-b)\log[J-b+c]}{2 a c},
}\\ \displaystyle{
d=0, c=0:\ K=\log I+\frac{1}{a} \log(J-b)-\frac{b}{a(J-b)}.
}
\end{array}
\right.
\label{eqAbelFirst} 
\end{eqnarray}
%
The first case mentioned in the list (\ref{eqAbelFirst}) corresponds to 
the solution (\ref{eq-gamma-power-half_alpha}).

Various assumptions extrapolating the three cases $b (a-1) d=0$, such as 
%
\begin{eqnarray}
& &
K=\log f(I)+k_+ \log[J+f_+(I)]+k_-\log[J+f_-(I)],
\end{eqnarray}
%
have failed to provide any new integrable case.

\section{Conclusion. Physical interpretation} 

Six closed form solutions $(\gamma(t), h(t))$ of the BDE
have been presented in this article:
(\ref{eqSol_gamma_constant}) and its sister solution for $\gamma=\gamma_0 (t/t_0)^{-\alpha}$, 
(\ref{eqSol-tpowern}), 
(\ref{eqSolpower-alpha_over-two}),
(\ref{eqSolPoint-sym-tn}),
(\ref{eqBDE-exponential-solution}).

If one excludes the unit of time $t_0$
and the unit of acceleration $\gamma_0$,
the number of their arbitrary parameters is respectively
two ($t_1$ and $K$),
one ($n$),
one ($c_1/c_2$),
two ($t_1$, $t_2$) and
one ($t_3$).

As far as we know, sister solution of (\ref{eqSol_gamma_constant}) and 
solutions (\ref{eqSolpower-alpha_over-two}), (\ref{eqSolPoint-sym-tn}) 
and (\ref{eqBDE-exponential-solution}) are new.

Moreover, there exist two values of $\gamma(t)$ allowing the BDE to reduce to 
a first order ODE of the type of Abel.
The only hope to integrate this Abel equation (\ref{eqAbel-IJ}) 
is to guess an integrating factor,
necessarily outside the classes (\ref{eqmu}) already examined by Abel, Liouville and 
Appell \cite{Appell89}.

In spite of its simplicity, solution (\ref{eqSol-tpowern}) has been found to play a key role. 
It has helped to exhibit two families of solutions for the spatial extension $h(t)$ of the mixing 
zone when 
the acceleration behaves asymptotically ($t \to +\infty$) like $\gamma(t) \sim t^n$ 
where $n$ is an arbitrary positive or negative exponent. 
For $n>0$, the growth of the acceleration is monotonic with time and $h$ 
increases like $h(t) \sim t^{n+2}$ for $t \to +\infty$. 
For $n<0$, the acceleration decreases asymptotically with time, and  
for large negative values of $n$, equation (\ref{eqBDE-RMI}) shows that the 
asymptotic solution does not depend anymore on $n$ but is given by $h_{RM}$ 
[see Eq.~(\ref{eqSolBDE-RMI-asymp})].
In this paper, we claim that $h(t) \sim t^{n+2}$ describes an acceleration-driven mixing, 
\textit {i.e.}~a mixing of Rayleigh--Taylor type while $h(t) \sim t^{1/(1+C)}$ corresponds 
to an acceleration decreasing too fast with time in order to be able to 
drive the evolution of the mixing zone and the solution of the BDE corresponds to a 
Richtmyer--Meskhov mixing.
The BDE exhibits therefore two leading behaviors and the way each of them arises 
is explained as follows: 
first, we notice that for $n \in [-2, -(1+2C)/(1+C)]$, solution (\ref{eqSol-tpowern}) 
is not physically valid since $h_0<0$ for $n \in ]-2, -(1+2C)/(1+C)[$ ~and $h_0$ diverges 
for the two values $n=-2$ and $n=-(1+2C)/(1+C)$.
Second, we show that for $n=-\alpha$ (this value is always below the range 
$[-2, -(1+2C)/(1+C)]$) and $n=-\alpha /2$ (this value is always in the range 
$[-2, -(1+2C)/(1+C)]$), the asymptotic solution is $h(t) \sim t^{1/(1+C)}$. 

As a consequence, we conclude that the threshold value for $n$ is $n_{th}=-(1+2C)/(1+C)$ 
(lower bound value (\ref{eqExposant-n-limite})) : 
for $n<n_{th}$, the acceleration decreases quickly with time and the system experiences the 
RMI with an asymptotic solution given by $h(t) \sim t^{1/(1+C)}$; for $n>n_{th}$, the system 
is driven by the RTI with the asymptotic solution $h(t) \sim t^{n+2}$. 

After submission of the present manuscript, we noticed that these two regimes had 
been already evidenced by Pandian, Swisher and Abarzhi 
under the name ``acceleration-driven mixing" and ``dissipation-driven mixing" for 
$n>n_{th}$ and $n<n_{th}$, respectively \cite{PSA}.

We notice that the RTI 
mixing occurs therefore also for $n_{th}<n<0$ although $\gamma(t)$ decreases with time.
Finally, for $n=n_{th}$, the two regimes collapse in a single one since $t^{n+2}=t^{1/(1+C)}$: 
the growth rates of the RTI and the RMI are the same. 
According to us, this claim is an important issue that would be worth being investigated 
through numerical simulations.

The solutions exhibited in this work provide an extension of the self-similar variable acceleration 
Rayleigh--Taylor (SSVART) flows studied by Llor \cite{LlorLPB,LlorLNP} where the author uses an 
acceleration $\gamma(t) \sim t^n$ for $t>0$ if $n>-2$, and $\gamma(t) \sim (-t)^n$ for $t<0$ if 
$n<-2$ (such flows have been also considered earlier by Neuvazhaev for $n>-1$ \cite{N83}).
In our case, the exponent $n=-2$ is not singular.
For this value of $n$, the direct application of (\ref{eqSol-tpowern}) leads to 
$h \sim t^{n+2} \sim t^0$ which of course is  
physically
wrong although the BDE is satisfied. 
Actually, this special value of $n$ satisfies $n<n_{th}$ and the correct asymptotic 
behavior is $h \sim t^{1/(1+C)}$. 

The case $n=0$ describes the standard RTI (\textit{i.e.} constant acceleration) and, 
as expected, (\ref{eqSol_asymp_gamma_constant}) is recovered from the particular 
solution (\ref{eqSol-tpowern}). For a constant acceleration, $h(t)$ is usually written 
as $h(t) \sim \alpha \gamma_0 t^2$ \cite{DimontePoF04} and the comparison with 
(\ref{eqSol_asymp_gamma_constant}) leads to the analytical value 
$\alpha = (B/2)/(1+2C)$. For $B=1$, the case $C=4$ leads to the smallest value 
$1/18 \approx 0.055$ for $\alpha$ in agreement with experiments whereas the 
value derived from numerical simulations is about twice smaller \cite{DimontePoF04}.

Finally, let us examine the case $n=-1$. 
This value is above $n_{th}$ and one expects the asymptotic solution 
$h \sim  h_0 (t/t_0)$ where the constant $h_0$ is given by (\ref{eqSol-tpowern}). 
Indeed, $\D ^2 h / \D t^2 = 0$ for this solution but the balance between $B \gamma (t)$ and 
$- (C/h)(dh/dt)^2$ is satisfied in the BDE.



\section*{Acknowledgments}

RC was partially supported by 
the Laboratoire de recherche conventionn\'e LRC-M\'eso.

\appendix

\section{Abel equation}
\label{sectionAbel}

We recall in this Appendix how to integrate an Abel equation.

The most general Abel equation 
(which can always be assumed to be of the first kind),
\begin{eqnarray}
& & -\frac{\D u}{\D x} + a_3(x) u^3 + a_2(x) u^2 + a_1(x) u +a_0(x)=0,
\label{eqAbelFirstkind}
\end{eqnarray}
is form-invariant under the mapping 
\begin{eqnarray}
& & (u,x) \to (U,X):\ x=F(X),\ u= P(X) U(X)+Q(X),\ F' P \not=0.
\label{eqAbelTransfo}
\end{eqnarray}

The classical method to decide whether it is integrable or not 
has been introduced by Roger Liouville \cite{RLiouvilleODE1}.
This first requires to compute its invariants under this mapping,
which are rational functions of the $a_j$'s and their derivatives.
The relative invariants of (\ref{eqAbelFirstkind}) have weight $2 m+1$, with $m$ a positive integer,
\begin{eqnarray}
& &
s_3=a_0 a_3^3
+\frac{1}{3}\left(\frac{2}{9} a_3^3 - a_1 a_2 a_3 + a_3 a_2' - a_2 a_3'\right),
\\ & &
s_{2 m+1}=a_3 s_{2m-1}' -(2 m-1)s_{2m-1}\left(a_3' + a_1 a_3  - \frac{1}{3} a_2^2\right),\
\end{eqnarray} 
and the absolute invariants $I_n$ are
\begin{eqnarray}
& &
I_1=\frac{s_5^3}  {s_3^5},\
I_2=\frac{s_3 s_7}{s_5^2},\
I_3=\frac{s_9}    {s_3^3},\ \dots
\end{eqnarray}
If $I_1$ is constant, all other absolute invariants are also constant
and there exists a mapping (\ref{eqAbelTransfo})
making the transformed equation separable.

If $I_1$ is not constant, integrating is equivalent to finding 
an integrating factor $\mu(u,x)$ to (\ref{eqAbelFirstkind}).
According to Abel, 
it is then more convenient to first put the equation under the form
\begin{eqnarray}
& &
u \frac{\D u}{\D x} + p(x) + q'(x) u=0, 
\end{eqnarray}
(whose advantage is to minimize the global degree in $u$ and $u'$),
then to make various assumptions for $\mu$, such as those considered by Abel,
\begin{eqnarray}
& & {\hskip -10.0truemm}
\log \mu=(b_1(x) u + b_0(x))^{-1},\ 
\mu=(b_1(x) u + b_0(x))^n,\ 
\mu=(u+b_1(x))^a (u+b_2(x))^b,\                       
\nonumber \\ & & {\hskip -10.0truemm}
\log \mu=b_3(x) u^3+ b_2(x) u^2 + b_1(x) u + b_0(x),\ 
\mu=\hbox{\cite[p.~215 Eq.~(48)]{ChebTerrab-Roche})}. 
\label{eqmu}
\end{eqnarray}

But the primary use of the invariants is to decide whether the equation 
belongs to the same equivalence class than one of the Abel ODEs previously integrated.
An outstanding presentation of the current achievements can be found in \cite{ChebTerrab-Roche},
where the authors reviewed all the cases listed in the books 
\cite{Kamke},
\cite{Murphy} 
\cite{PolyaninZaitsev},
added a few ones
and ordered them in different classes of equivalence \textit{modulo} (\ref{eqAbelTransfo}).

Let us mention a different approach \cite{Pana2011} to try to integrate Abel ODEs. 
The authors split the single ODE into a system made of one Riccati ODE and another condition.
If the Riccati ODE can be integrated, then this may lead to the integration of the Abel ODE.
However, this method has not yet produced new integrated Abel equations.

%
%
%


\vfill \eject 
\end{document}